\documentclass[aps,prd,onecolumn,groupedaddress,showpacs,showkeys, nofootinbib,amssymb]{revtex4}
\usepackage[T1]{fontenc}
\usepackage[latin1]{inputenc}
\usepackage{graphicx}
\usepackage[english]{babel}
\usepackage{bm}
\usepackage{amsmath}
\usepackage{amssymb}
\usepackage{amsfonts}
\usepackage{epsfig}
\usepackage{colordvi}
\usepackage{color}

\begin{document}

\title{Probing the physical and mathematical structure of  $f(R)$ gravity by PSR  $J0348+0432$}

\author{Mariafelicia De Laurentis\footnote{mfdelaurentis@tspu.edu.ru}}

\address{Tomsk State Pedagogical University, Tomsk, ul. Kievskaya 60, 634061 Russian Federation}

\author{Ivan  De Martino}

\address{Departamento de  Fisica Teorica, Universidad de Salamanca, 37008 Salamanca, Spain.\\
 INFN Sezione  di Napoli, Compl. Univ. di
Monte S. Angelo, Edificio G, Via Cinthia, I-80126, Napoli, Italy.}

\begin{abstract}
There are several approaches to extend General Relativity in order to explain the phenomena related 
to the Dark Matter and Dark Energy.  These theories, generally called Extended Theories of Gravity, can be tested using observations coming 
from relativistic binary systems as  PSR $J0348+0432$.  Using a class of analytical $f(R)$-theories, 
one can construct the first time derivative of orbital period of the binary systems 
starting from a quadrupolar gravitational emission.  Our aim is to
set boundaries on the parameters of the theory in order to understand if they are ruled out, or not, by the observations 
on PSR $J0348+0432$.  Finally, we have computed an upper limit on the graviton mass showing that agree with constraint coming from
other observations. 
\end{abstract}

\maketitle

\section{Introduction}\label{sec:intro}

Einstein's theory of gravity, known as General Relativity (GR), is a well description of gravitational
phenomena at astrophysical and cosmological scales.  The theoretical predictions are in good agreement with  
the measurements in particular at the Solar System scale \cite{will}.  Even if,  the local gravity is very well described 
by GR, there are several
observations at astrophysical and cosmological scales for which we need to add extra ingredients in the amount 
of matter and energy densities of the Universe to fit the data.  Observations based on Supernovae 
Type Ia (SNeIa),  Cosmic Microwave Background (CMB) temperature anisotropies, Baryon Acoustic Oscillations (BAO)
and other observables, pointed out that the Universe is in a period of accelerated expansion favoring the concordance
$\Lambda$CDM model \cite{astier, bao:blake, PLANCKXVI2013, riess}.  In this model, we need to input an extra component, 
known as Dark Energy (DE), that have the effect of a fluid with a negative pressure, to explain the acceleration of the 
Universe at cosmological scales \cite{DEreview}.  Furthermore, we also need to input an extra contribution in the amount of matter, known as Dark Matter (DM)
to explain the clustering of  the Large Scale Structure.  There are observational evidences point out that the baryonic 
matter is not enough to explain phenomena at galaxy and cluster scales, like rotation curves, gravitational lensing, 
cluster profiles and others \cite{DMreview}.  However, we do not know the nature, at particle level, of this two
dark ingredients, even if there are many explanations in literature: quintessence, string theory, holographic principle 
are examples of DE models, while Weakly Interacting Particles (WIMPS), Axions, or Massive Compact Halo Objects 
(MACHOs) was been proposed like candidates for DM.  As an alternative, instead to introduce two unknown 
components to explain the dynamics and the evolution of Universe at all scales, it is possible change its the geometrical 
description.  In this framework the Extend Theories of Gravity  (ETGs) are been developed \cite{PRnostro,PRsergei,faraoni,olmo}. 
By introducing the higher curvature terms in the Lagrangian allow us to reconstruct the DE and DM 
phenomenology at all scales, from planetary dynamics, to flat rotation curves of spiral galaxies 
 and the velocity dispersion of ellipticals, and from cluster of galaxies
until the dynamics of the Universe as a whole \cite{annalen,ivan_cluster}.

However, to constraint and validate ETGs
could be extremely important take also into account the gravitational waves (GWs) emission.  In GR, the field equations linearized around a Minkowskian 
metric, show that small perturbations of the metric propagate following a wave equation \cite{Maggiore, gravitation}.  
All astrophysical systems that emit GWs  are very well described in the framework of GR, where the gravitational
interaction is mediate by a massless boson known as graviton.  But, the further degrees of freedom that come out 
considering ETGs lead to have also massive gravitational modes.

In particular, from the power spectrum of weak lensing one can estimate an upper limit of $7\times 10^{-32}$
eV for the graviton mass. 
Then, from clusters of galaxies,  it is possible  to obtain an upper limit of $2\times 10^{-29} h_{0}$ eV, where
$h_0$ is the Hubble constant in units of 100\,km\,s$^{-1}$\,Mpc$^{-1}$.  
Finally, studying the emission
 of GWs from the binary systems, one can infer an upper limit of graviton mass  of $7.6\times
10^{-20}$ eV  \cite{will}. 

Since the mass of graviton have also effect of the waveform and on
the polarization modes,  the detection of GWs will point out that GR is validated or  that it must be extended \cite{PRnostro}. 
At moment, the most important tool, regarding to GWs, to constrain both theories of gravity is related to
the gravitational emission from binary systems of White Dwarfs (WDs), Neutron Stars (NS) and Black Holes (BHs) 
\cite{Maggiore}. 
Timing data analysis on the well-known binary pulsar B$1913+16$ have confirmed that the energy loss by the system can 
be explained with the emission of GWs.  Let's point out that those type of systems, due to the high precision of 
the mass estimation, and also for the strong field gravity regime in which they are (PSR B$1913+16$
or  PSR J$0348+0432$ are two examples of system with very precise measuraments), 
represent a very good laboratory to test theories of gravity using Post-Keplerian 
parameters \cite{ ht75a, wnt10,Damour-Farese, af+13}.  Let's start to consider 
a generic class of ETGs, called $f(R)$-theories, where we replace the Einstein-Hilbert Lagrangian with a more general 
function of the Ricci curvature.  For an analytic $f(R)$, it is possible evaluate 
the first time derivatives of the orbital period for a binary system and, comparing the theoretical estimation 
with the observed one \cite{nostromnras}, we put bounds on theory parameters and then we compute an upper limit on the graviton mass. 

The outline of the paper is the following: in Sec. \ref{due} we briefly introduce the theoretical framework in which we describe 
binary systems computing the energy lost through GWs emission.  In Sec. \ref{tre} we summarize how to compute the first time 
derivative on the orbital period for a binary system in $f(R)$-gravity.  Furthermore, we
test how well the PSR $J0348+0432$ can constrain the first time derivative of  
the orbital period in $f(R)$-theories of gravity and, estimate the bounds on graviton mass.  
Finally, in Sec. \ref{quattro} we give our conclusions and remarks.

\section{Theoretical framework: $f(R)$-gravity}
\label{due}

 In ETGs, the field equations are computed,  extending the Hilbert-Einstein Lagrangian with adding of higher-order curvature invariants and minimally  or non-minimally coupled scalar fields.  The simplest way is to consider a more general function of the curvature $f(R)$ \cite{PRnostro}. 
 In this case, the field equations have the following form 
(looking for major details at \cite{PRnostro,PRsergei,faraoni,olmo}) 
\begin{eqnarray}\label{fe1}
f'(R)R_{\mu\nu}-\frac{f(R)}{2}\,g_{\mu\nu}-f'(R)_{;\mu\nu}+g_{\mu\nu}\Box_g f'(R)\,=\,\frac{\mathcal{X}}{2}T_{\mu\nu}\,,
\end{eqnarray}
\begin{eqnarray}\label{TrHOEQ}
3\Box f'(R)+f'(R)R-2f(R)\,=\,\frac{\mathcal{X}}{2}T\,,
\end{eqnarray}
where ${\displaystyle T_{\mu\nu}=\frac{-2}{\sqrt{-g}}\frac{\delta(\sqrt{-g}\mathcal{L}_m)}{\delta
g^{\mu\nu}}}$ represents  the energy momentum tensor of matter ($T$ is the
trace), ${\displaystyle \mathcal{X}=\frac{16\pi G}{c^4}}$ is the coupling, ${\displaystyle f'(R)=\frac{df(R)}{dR}}$,
$\Box_g={{}_{;\sigma}}^{;\sigma}$, and $\Box={{}_{,\sigma}}^{,\sigma}$ indicates the d'Alembert operator.  
The adopted signature is $(+,-,-,-)$, and we indicate  the partial derivative with "$,$" and the covariant derivative with 
respect to the $g_{\mu\nu}$ is indicated by " $;$" ; all Greek indices go from $0$ to $3$ and Latin indices go from $1$ 
to $3$; $g$ is the metric determinant.   
In a more general approach, we don't fix the form of the $f(R)$ Lagrangian but we just assume that it is Taylor 
expandable in term of the Ricci scalar\footnote{For convenience we will use $f$ instead of $f(R)$.  
All considerations are developed here in metric formalism.  From now on we assume physical  units $G=c= 1$. }
in order to compute the Minkowskian limit of the theory  \cite{quadrupolo,landau,nostromnras},

\begin{eqnarray}\label{sertay}
f(R)=\sum_{n}\frac{f^n(R_0)}{n!}(R-R_0)^n\simeq
f_0+f'_0R+\frac{f''_0}{2}R^2+... \,. 
\end{eqnarray}

In this contest,  using theoretical approach given in \cite{landau}, 
the total average flux of energy emitted in GWs has the following form  
\begin{eqnarray}
\underbrace{\left\langle \frac{dE}{dt}\right\rangle}_{(total)}&=& \frac{G}{60}\left\langle \underbrace{ f'_0  \left({ \dddot Q}^{ij}{ \dddot Q}_{ij}\right)}_{GR}
-\underbrace{f''_0 \left({ \ddddot Q}^{ij}{ \ddddot Q}_{ij}\right)}_{f(R)}
 \right\rangle\,. 
 \label{energy}\end{eqnarray}
 The ratio of $f'_0$ and $f''_0$ defines an effective mass 
 $m_{g}^2\,=\,-\frac{f'_0}{3f''_0}$ related to massive modes in GWs \cite{greci}. 
Until now, the only modes of gravitational waves that have been investigated are the massless ones, 
this exclude the possibility to observe further gravitational waveform coming out from massive terms,
although, tests in this direction are already done on stochastic background of GWs  \cite{bellucci,stoca1,stoca3,stoca2}. 

From the equations of conservation we know that the gravitational radiation, predicted by GR, is proportional 
to the third derivative of the quadrupole momentum, while the terms due to the monopole and the dipole momentum are
zero.  For a discussion on gravitational emission see \cite{Will,Damour,jetzer,PRnostro}.
In $f(R)$-gravity we find  a different situation.  
The fact that we have extended the theory
of gravity can be reflected by a mass of graviton not equal to zero.  As well known, in  ETG it is  possible to obtain
massive gravitons in a natural way \cite{greci}. The main feature is that higher-order terms or induced scalar fields in the Lagrangian, 
give rise to massless, massive spin-$2$ gravitons and massive spin-$0$ gravitons.  
Such gravitational modes results in $6$ polarizations,
according to the Riemann Theorem stating that in a given $n$-dimensional space, $n(n-1)/2$
degrees of freedom are possible.  
The fact that 6 polarization states emerge is in agreement with the possible allowed polarizations of spin-2 field 
\cite{van}.  In fact, the spin degenerations is
\begin{itemize}
\item $d = (2s+1), \quad  m_g\neq 0 \Longrightarrow  s=2,\quad d\,=\,5 $    
\item $d = 2s,   \quad m_g = 0  \Longrightarrow  s = 1, \quad d\,=\,2     $
\item  $d = (2s+1),  \quad    m_g \neq 0  \Longrightarrow    s = 0,\quad d\,=\,1$               
\end{itemize}
The massive spin-$2$ gravitational states, usually are ghost particles.   
The role of massive gravitons result relevant also in the case which
we want define a cutoff mass at TeV scale.  This limits  allow us, both to circumvent
the hierarchy problem that the detection of the Higgs boson. 
For example, in such a case, the Standard Model of particles should
be confirmed without recurring to perturbative, renormalizable
schemes involving new particles \cite{basini,luca,giulia}.
Furthermore,  ETGs in post Newtonian regime gives rise at the Yukawa-like correction to the Newtonian potential.  
These corrections are dependent by  a characteristic length of self-gravitating structures that is connected to massive graviton modes. 
Upper limits on graviton mass  come out when ones try to solve the connection between  masses and sizes of self-gravitating structures without invoking huge amounts of DM.

\section{The first time derivative of the orbital period of a binary system: $f(R)$ parameters and bounds on graviton mass}
\label{tre}
Following the scheme that is generally used to compute GWs emission \cite{Maggiore}, and assuming
a Keplerian motion of the stars in the binary system, we can define $m_p$ as the pulsar mass, 
$m_c$ as the companion mass, and $\displaystyle{\mu=\frac{m_c m_p}{m_c+m_p}}$ as the reduced mass.  
The motion is reduced at the $(x-y)$-plane, so that averaging on the orbital period, $P_b$, and using the 
eqs.  \eqref{energy}, we get the first time derivative of the orbital period \cite{nostromnras}
\begin{align}\label{periodvariation}
&\dot{P}_b =  - \dfrac{3}{{20}}{\left( {\dfrac{P_b}{{2\pi }}} \right)^{ - \frac{5}{3}}}\dfrac{{\mu {G^{\frac{5}{3}}}{{({m_c} + {m_p})}^{\frac{2}{3}}}}}{{{c^5}{{(1 - {\epsilon^2})}^{\frac{7}{2}}}}}
 \times \left[ {{f'_0}\left( {37{\epsilon ^4} + 292{\epsilon ^2} + 96} \right) - \dfrac{{{f''_0}{\pi ^2}{T^{ - 1}}}}{{2{{(1 + {\epsilon^2})}^3}}} \times } \right. \nonumber\\
&\left.  {\times \left( {891{\epsilon ^8} + 28016{\epsilon ^6} + 82736{\epsilon ^4} + 43520{\epsilon ^2} + 3072} \right)} \right].  
\end{align}
where $\epsilon$ is the eccentricity of the orbit, $G$ is the Newtonian gravitational constant, and the quantities
${f'_0}$ and ${f''_0}$ are the ones that we need to constrain.  Once we have a theoretical prediction of $\dot{P}_b$ in 
$f(R)$-theories  we can, according with the prescription given in \cite{fs+02}, compute an upper limit for the graviton 
mass.

\subsection{Application to PSR J0348+0432}

The binary system PSR $J0348+0432$,  recently studied by Antoniadis et al.  (2013) \cite{af+13}, gives a new possibility 
to understand which range of ETG's parameters is allowed.  It is a binary system composed by a pulsar spinning at 
$39$ ms with mass $2.01\pm0.04$ $M_\odot$ ($m_p$), and a White Dwarf (WD) companion with mass 
$0.172\pm0.003\, M_{\odot}$ ($m_c$).  The orbital period of the system is  $P_b=0.102$ (days), and the eccentricity 
$e=2.36008\times10^{-6}$.  For a correct estimation of the observed orbital decay, 
it was considered several kinematic effects that have to be subtracted by
variation of the orbital period.  The first one is the Shklovskii effect \cite{shk70}, that is an expression of
the effect due to the proper motion of the the binary system.  Its estimation for this system is 
\begin{equation}
\dot{P}_{\rm b}^{\rm Shk} = P_{\rm b} \frac{\mu^2 d}{c} =
0.0129^{+0.0025}_{-0.0021} \times 10^{-13}. 
\end{equation}

Another effect is due to the difference of Galactic accelerations between 
the binary system and the Solar System
\begin{equation}
\dot{P}_{\rm b}^{\rm Acc} = P_{\rm b} \frac{a_c}{c} =
0.0037^{+0.0006}_{-0.0005} \times 10^{-13}. 
\end{equation}

The last term is due to a possible variation of the gravitational 
constant $\dot{G}$:
\begin{equation}
  \dot{P}_{\rm b}^{\dot{G}} 
  = - 2 P_{\rm b} \frac{\dot{G}}{G} 
  = (0.0003\pm 0.0018) \times 10^{-13}. 
\end{equation} 

Finally, the observed value of the first time derivative of the orbital period is 
$\dot{P}_b=(-2. 73 \pm 0. 45) \times 10^{-13}$.  All those terms are quoted in  \cite{af+13}.  

As pointed out in 
\cite{nostromnras}, the ETGs are not ruled out when the orbital parameters of the binary systems 
are very well estimated, and the range of the $f(R)$ parameters is $0.04 \leq| f''_0|\leq 38$.

\begin{figure}
\centering
\includegraphics[scale=0.75]{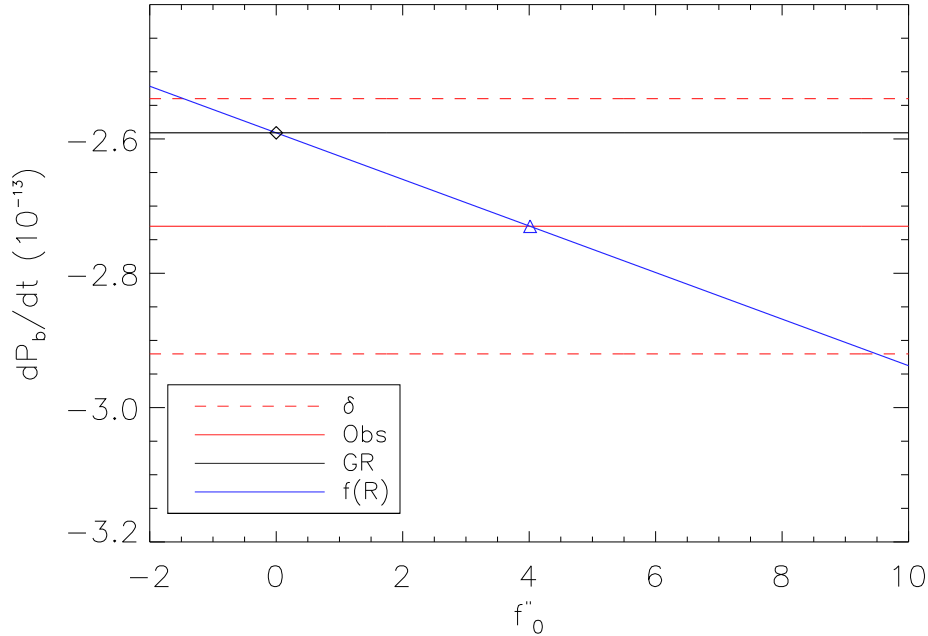}
\caption {It is shown the result of our numerical analysis on the  binary system PSR$J0348+0432$. 
We use the following notation: the black line shows the behavior of the GR prediction for the first derivative of the orbital period for 
a binary system; the red line represents the observed orbital period variation ${\dot P_{b_{Obs}}}$ and its errors,
in particular the dashed lines show the experimental error on the observation; the blue line shows the $f(R)$-theory prediction
as computed in eq.  \eqref{periodvariation}. }\label{fig:fig1}
\end{figure}

In Fig. \ref{fig:fig1}, we report the result of our numerical analysis on the  binary system PSR $J0348+0432$. 
We use the following notation: the black line shows the behavior of the GR prediction for the first derivative 
of the orbital period for the binary system; the red line represents the observed orbital period variation 
${\dot P_{b_{Obs}}}$ and its errors, in particular the dashed lines show the experimental error on the observation; 
the blue line shows the $f(R)$-theory prediction as computed in eq.  \eqref{periodvariation}. 
Moreover, it is possible to see from Fig. \ref{fig:fig1} that  the GR value of ${\dot P_{GR}}$ 
is recovered for $f''_0= 0$ (diamond black).  The value of the second derivative of the gravitational Lagrangian 
for which ETGs are able to explain 
 the observed period variation is $f''_0=4.02\pm5.48$ (blue triangle).  
This value is comparable with the range given in \cite{nostromnras}.  
Using the range of the $f(R)$ parameters allowed, we are interested to compute an upper limit for the graviton mass
to understand if this agree with other estimation.  According with the prescription given in \cite{fs+02}
we get the following upper limit
\begin{equation}
 m_g < 5.95\times10^{-20} {\rm{eV}}/{c^2}, 
\end{equation}
that is comparable with the constrain coming out from $B1913+16$, and is also comparable with the one obtained analyzing 
the Brans-Dicke theory \cite{af+13}.

\section{Discussion}
\label{quattro}

Several shortcomings coming out from astrophysical and cosmological observations, and theoretical problem
related to have a full quantum description of space-time, suggest to extend the  GR
to overcome them.  Approaches based on extension or corrections of GR  are in agreement with several 
observables at astrophysical and cosmological scales, explaining the dynamical effect related to the DM, and
also the observed acceleration of the Universe. 

In the post-Minkowskian limit of ETGs provide an accurate description of 
the problem of gravitational radiation.  In this approach, we found extra polarizations modes 
of the GWs signal with respect to the plus (+) and cross ($\times$) polarizations predicted by GR \cite{PRnostro}. 
Indeed, the theories allows massive and ghosts modes that in principle could be detected using ground-based 
interferometric detectors, like VIRGO and LIGO \cite{VIRGO,LIGO}, and the future space interferometric detector, LISA \cite{LISA}
including the additional polarization modes.  Furthermore, results coming out from the timing array analysis on 
binary pulsars systems like PSR $B1913+16$  and from other binary systems, can provide a very useful tool to 
investigate the  the viability of ETGs, like the $f(R)$-gravity theories.  
For that purpose, it is possible to develop a quadrupolar formalism of the gravitaional radiation using 
analytic $f(R)$-models, and showing that expertimental bounds on the first derivative of the orbital period
of the binary system, that are provided by  PSR $J0348+0432$, do not ruled out the 
$f(R)$-gravity \cite{quadrupolo, nostromnras}.   Furthermore, the upper limit on graviton mass, that we have computed
basing our calculation on the experimental constraint of the orbital parameters, and on the theoretical prediction
of the $f(R)$ model, is comparable with the other constrains that come from clusters, galaxies etc \cite{af+13}.  
This result, together with other results from observations in the Solar System and Cosmology, indicates that the 
study and the possible detection of massive modes of GWs could be the real discriminating 
between the theory of GR and its extensions.

\section*{Acknowledgments}

This work has been partially funded by Ministerio de Educacion y Ciencia (Spain), grants FIS2012-30926 and N FIS2009-07238. 
 MDL and ID  acknowledge the support of INFN Sez.  di Napoli (Iniziativa
Specifica TEONGRAV).

\end{document}